**ORIGINAL ARTICLE**

**Outflow boundary conditions for three-dimensional simulations of non-periodic blood flow and pressure fields in deformable arteries**

I. E. Vignon-Clementel[1*], C. A. Figueroa[2], K. E. Jansen[6], C. A. Taylor[2,3,4,5]

[1] *INRIA Paris-Rocquencourt, Le Chesnay, France; Departments of* [2]*Bioengineering,* [3]*Mechanical Engineering,* [4]*Pediatrics, and* [5]*Surgery, Stanford University, Stanford, CA 94305, USA and* [6]*Aerospace Engineering Sciences, University of Colorado in Boulder, Boulder, CO 80309, USA*

*Corresponding author. Email irene.vignon-clementel@inria.fr

Abstract

The simulation of blood flow and pressure in arteries requires outflow boundary conditions that incorporate models of downstream domains. We previously described a coupled multidomain method to couple analytical models of the downstream domains with three-dimensional numerical models of the upstream vasculature. This prior work either included pure-resistance boundary conditions or impedance boundary conditions based on assumed periodicity of the solution. However, flow and pressure in arteries are not necessarily periodic in time due to heart-rate variability, respiration, complex, transitional flow or acute physiological changes. We present herein an approach for prescribing lumped parameter outflow boundary conditions that accommodate transient phenomena. We have applied this method to compute hemodynamic quantities in different physiologically relevant cardiovascular models, including patient-specific examples, to study non-periodic flow phenomena often observed in normal subjects and in patients with acquired or congenital cardiovascular disease. The relevance of using boundary conditions that accommodate transient phenomena compared to boundary conditions that assume periodicity of the solution is discussed.

# 1   Introduction

The choice of outflow boundary conditions can have a significant influence on velocity and pressure fields in three-dimensional simulations of blood flow. While prescribed velocity or pressure outflow boundary conditions are typically applied, this approach is inappropriate when modeling wave propagation phenomena in human arteries for a number of reasons. First, obtaining such data - especially time-varying data - for each outlet is impractical. Second, even if time-varying flow or pressure were acquired simultaneously for each outlet, it would be very difficult to synchronize these waveforms in a manner consistent with the wave propagation linked to the wall properties of the numerical domain. Indeed, current simulation capabilities include either rigid wall assumptions or limited information about the vessel wall properties (elastic modulus, thickness and their spatial variations). Finally, prescribing time-varying flow or pressure is not relevant for treatment planning applications where the quantity of blood flow exiting branch vessels and the distribution of pressure are unknown and part of the desired solution.

   A commonly used boundary condition type that does not require the specification of either flow rate or pressure is the resistance boundary condition. However, this condition severely impacts wave propagation phenomena, since it forces the flow and pressure waves to be in phase and it can generate aberrant pressure values in situations of flow reversal (Vignon-Clementel et al. 2006, Vignon and Taylor 2004). A better strategy is to use one-dimensional methods to model the downstream vessels and provide boundary conditions for the more computationally-intensive, three-dimensional methods modeling the major arteries (Formaggia et al. 2001, Lagana et al. 2002, Taylor and Hughes 1998, Urquiza et al. 2006, Vignon-Clementel et al. 2006). However, solving the transient non-linear one-dimensional equations of blood flow in the millions of downstream vessels is an intractable problem and therefore linearized one-dimensional models are needed. These simplified linear methods usually assume periodicity of the solution (Brown 1996, Olufsen 1999,

Olufsen et al. 2000, Spilker et al. 2007, Steele et al. 2007, Vignon-Clementel et al. 2006). Yet, blood flow and pressure in arteries are not always periodic in time due to heart-rate variability, respiration, acute physiological changes, or transitional or turbulent flow (DeGroff et al. 2005, Nichols and O'Rourke 2005, Otsuka et al. 1997, Sherwin and Blackburn 2005). Thus, in the present work, zero-dimensional (lumped) models are directly coupled to the three-dimensional equations of blood flow and vessel wall dynamics.

A lumped parameter model is a dynamic description of the physics neglecting the spatial variation of its parameters and variables. If the model is distributed, these parameters and variables are assumed to be uniform in each spatial compartment. Therefore, a lumped parameter model is described by a set of ordinary differential equations representing the dynamics in time of the variables in each compartment. Several groups have successfully coupled the fully three-dimensional models for pulsatile blood flow to either resistances (Bove et al. 2003, Guadagni et al. 2001, Migliavacca et al. 2003, Taylor et al. 1999) or more sophisticated lumped models with a Windkessel model (Torii et al. 2006) or an extensive model of the whole vasculature (Lagana et al. 2002, Lagana et al. 2005). Different strategies for the coupling of the three-dimensional equations with lumped models have been presented in (Quarteroni et al. 2001). The well-posedness analysis of this coupling has been studied in (Quarteroni and Veneziani 2003). However, in all the above-mentioned articles, this coupling has been performed iteratively, which can lead to stability and convergence issues, and has been generally applied to geometries with one or two outlets, rigid walls and low resistances (as seen in the pulmonary vasculature). We note here the work of Urquiza et al. (Urquiza et al. 2006) on a monolithic approach to couple the 3D fluid-structure interaction with 1D models that are themselves coupled to 0D models, presenting a unified variational approach for 3D-1D coupling (Blanco et al. 2007). Quasi- and aperiodic phenomena have been studied in a one-dimensional collapsible tubes (Bertram 1995, Jensen 1992). Coupling one-dimensional to lumped models, Stergiopulos et al. developed a general method to study nonlinear pressure and flow wave propagation that can be used to model non-periodic and transient phenomena such as heart-rate changes, stress, Valsalva maneuver, etc. (Stergiopulos et al. 1993).

The work presented herein exhibits several differences with prior work. First, we employ the three-dimensional elastodynamics equations to describe the vessel wall structural response to blood flow and pressure (Figueroa et al. 2006). Second, we demonstrate the capabilities of this method on patient-specific multi-branched geometries. Third, a coupled 3D-0D approach has been used to study dynamic changes due to heart rate variability, respiratory effects, or non-periodic flow phenomena that may arise from congenital and acquired vascular disease.

In this work, we extend the "coupled multidomain method" derived to couple analytical models of the downstream domains with three-dimensional numerical models of the upstream vasculature (Vignon-Clementel et al. 2006) to include boundary conditions that accommodate non-periodic phenomena using lumped parameter (e.g. Windkessel) outflow boundary conditions. The coupling of blood flow and pressure with the vessel wall dynamics is presented in (Figueroa et al. 2006). Note that this prior work either included pure-resistance boundary conditions, which do not require periodicity of the solution but cannot be used to accurately model flow and pressure waveforms (Vignon-Clementel et al. 2006, Vignon and Taylor 2004), or impedance boundary conditions based on assumed periodicity of the solution. In this paper, we have applied this method to compute three-dimensional pulsatile flow and pressure in a simple model of the common carotid artery with pulsatility changes, in a patient specific carotid bifurcation with a severe stenosis that causes transitional flow (acquired cardiovascular disease) with and without inflow periodic variability and in a patient-specific Glenn geometry with complex multiple branches and pulsatility changes (congenital cardiovascular disease). These cases have in common some non-periodicity in the flow and pressure. However they represent a broad class of applications and we have thus introduced their biomechanical and clinical specificities in their respective results subsections. The discussion section addresses some limitations, verification and validation aspects related to boundary conditions. In addition, the results imposing fully transient or periodicity based boundary conditions are compared and discussed.

## 2   Methods

In this section, we present a summary of the methods we developed to model the influence of the "downstream" vasculature on blood flow and vessel wall dynamics in a three-dimensional computational domain. We then present the extension of these methods to lumped parameter (e.g. Windkessel) boundary conditions.

### 2.1   *Variational formulation for blood flow, pressure and wall deformation*

The coupled multidomain method is presented in (Vignon-Clementel et al. 2006) in the context of rigid walls. The strategy to model the interactions between blood and the artery walls is derived in (Figueroa et al. 2006). Here we summarize the main steps of the combination of the two methods:

We consider the Navier-Stokes equations to represent the three-dimensional motion of blood in a domain $\Omega$ over time zero to $T$. We can formulate the balance of mass and momentum as follows:

Given $\vec{f}:\Omega\times(0,T)\rightarrow\Re^{3}$, find $\vec{v}(\vec{x},t)$ and $p(\vec{x},t)$ $\forall\vec{x}\in\Omega$, $\forall t\in(0,T)$ such that :

$$\rho\vec{v}_{,t}+\rho(\nabla\vec{v})\vec{v}=div(\underline{\tau})+\vec{f}$$
$$div(\vec{v})=0 \qquad\qquad (1)$$
$$\underline{\tau}=-p\underline{I}+\mu(\nabla\vec{v}+\nabla\vec{v}^{T})$$

The primary variables are the blood velocity $\vec{v}=(v_{x},v_{y},v_{z})$ and the pressure $p$. The blood density is given by $\rho$ (assumed constant), the external force by $\vec{f}$ and the dynamic blood viscosity by $\mu$ (assumed constant).

We consider the elastodynamics equations to describe the motion of the arterial wall represented by a domain $\Omega^{s}$:

Given $\vec{b}^{s}:\Omega^{s}\times(0,T)\rightarrow\Re^{3}$, find $\vec{u}(\vec{x},t)$ $\forall\vec{x}\in\Omega^{s}$, $\forall t\in(0,T)$ such that:

$$\rho^{s}\vec{u}_{,tt}=\nabla\cdot\underline{\sigma}^{s}+\vec{b}^{s}$$
$$\underline{\sigma}^{s}=\underline{C}:\underline{\varepsilon} \qquad\qquad (2)$$
$$\underline{\varepsilon}=\frac{1}{2}\left(\nabla\vec{u}+\nabla\vec{u}^{T}\right)$$

The primary variable is the vessel wall displacement $\vec{u}$. The wall density is given by $\rho^{s}$ (assumed constant), the external body force by $\vec{b}^{s}$ and the linearized material behavior by $\underline{C}$. We characterize the structural response of the arterial wall using an enhanced thin membrane model whereby the domain $\Omega_{s}$ is represented by the lateral boundary of the blood domain $\Gamma_{s}$ and the wall thickness $\zeta$ (i.e., $\Omega^{s}\equiv\Gamma_{s}\times\zeta$), which is considered here as a parameter. On this boundary $\Gamma_{s}$, we enforce the kinematic compatibility condition so that the blood and vessel wall have the same velocity:

$$\vec{v}=\vec{u}_{,t} \qquad\qquad (3)$$

The dynamic compatibility condition is satisfied by defining a wall body force $\vec{b}^{s}$ from the wall surface traction $\vec{t}^{s}$, which is equal to the opposite of the traction $\vec{t}^{f}=\underline{\tau}\cdot\vec{n}$ felt by the blood on the boundary $\Gamma_{s}$. In particular, using a thin-wall assumption (see (Figueroa et al. 2006)) we have

$$\vec{b}^s = \frac{\vec{t}^s}{\zeta} = -\frac{\vec{t}^f}{\zeta} \tag{4}$$

We assume that the edges of the vessel wall domain are fixed. Considering this, a single variational form is derived for the fluid domain $\Omega$, incorporating the influence of the vessel wall on the boundary $\Gamma_s$. Appropriate initial and boundary conditions are needed to complete these sets of equations (for more details, see (Figueroa et al. 2006)).

A disjoint decomposition of the variables is performed in $\Omega = \hat{\Omega} \cup \Omega'$ separating the computational domain $\hat{\Omega}$ (and its vessel wall boundary $\hat{\Gamma}_s$) from the downstream domains $\Omega'$. The interface that separates these domains is defined as $\Gamma_B$. The variational formulation can be rewritten in terms of the variables in the computational domain $\hat{\Omega}$ only. The resulting weak form of the fluid-solid interaction problem with the multidomain coupling is:

Given the material parameters defined above, body forces $\vec{f}:\Omega \times (0,T) \to \Re^3$ and prescribed velocities $\vec{g}:\Gamma_g \times (0,T) \to \Re^3$, find the velocity $\vec{v}$ and the pressure $p$ such that for every test function $\vec{w}$ and $q$

$$\int_{\hat{\Omega}} \hat{\vec{w}} \cdot \left( \rho \hat{\vec{v}}_{,t} + \rho \hat{\vec{v}} \cdot \nabla \hat{\vec{v}} - \vec{f} \right) + \nabla \hat{\vec{w}} : \left( -\hat{p}\underline{I} + \hat{\underline{\tau}} \right) d\vec{x}$$

$$+\zeta \int_{\hat{\Gamma}_s} \left\{ \hat{\vec{w}} \cdot \rho^s \hat{\vec{v}}_{,t} + \nabla \hat{\vec{w}} : \underline{\sigma}^s \left( \hat{\vec{u}} \right) \right\} ds \boxed{-\int_{\Gamma_B} \hat{\vec{w}} \cdot \left( \underline{M}_m (\hat{\vec{v}}, \hat{p}) + \underline{H}_m \right) \cdot \hat{\vec{n}} ds} \tag{5}$$

$$-\int_{\hat{\Omega}} \nabla \hat{q} \cdot \hat{\vec{v}} d\vec{x} + \int_{\hat{\Gamma}} \hat{q} \hat{\vec{v}} \cdot \hat{\vec{n}} ds + \boxed{\int_{\Gamma_B} \hat{q} \left( \vec{M}_c (\hat{\vec{v}}, \hat{p}) + \vec{H}_c \right) \cdot \hat{\vec{n}} ds} = 0$$

Note that influence of the downstream domains naturally appears in the boundary fluxes (boxed terms in (5)) on $\Gamma_B$ (see (Vignon-Clementel et al. 2006)). The momentum $\underline{M}_m, \underline{H}_m$ and continuity $\vec{M}_c, \vec{H}_c$ operators depend on the model chosen to represent the hemodynamics conditions in the downstream domain $\Omega'$ (lumped models, one-dimensional equations of blood flow, pressure and vessel wall deformation, etc.). As a result of the continuity of stress and mass fluxes through the interface $\Gamma_B$, these operators act solely on the unknowns of the numerical domain $\hat{\vec{v}}$ and $\hat{p}$. In this paper, we focus on cases where at the inlet a velocity profile is prescribed as a function of time. Downstream domains are discussed in the following section.

### 2.2 Time-varying outlet boundary condition

There are several techniques to model the effects of the downstream vasculature in the upstream computational domain. The resistance and one-dimensional impedance boundary conditions have already been presented in (Vignon-Clementel et al. 2006), but here we explore a class of boundary conditions that can accommodate transient phenomena. We present one of the simplest models to demonstrate the capabilities of this approach, but any ordinary differential equation relating flow and pressure can be used in a similar way. The RCR (also known as "Windkessel") model is an electric circuit analog that has a proximal resistance $R$ in series with a parallel arrangement of a capacitance $C$ and a distal resistance $R_d$. The Windkessel model was originally derived by the German physiologist Otto Frank in an article published in 1899 (Frank 1899). A downstream pressure $P_d$ varying as a function of time can be used to represent the terminal pressure (Figure 1). However, in many cases, this terminal pressure is assumed to be zero.

At time $t$, the pressure $P$ is related to the flow rate $Q$ by the following relationship (assuming the simulation starts at $t = 0$):

$$P(t) = \left[ P(0) - RQ(0) - P_d(0) \right] e^{-t/\tau} + P_d(t) + RQ + \int_0^t \frac{e^{-(t-\tilde{t})/\tau}}{C} Q(\tilde{t}) d\tilde{t} \tag{6}$$

$$\tau = R_d C$$

The time constant $\tau$ describes how fast the system responds to a change in the input function. As can be seen in equation (6), pressure at time $t$ is related to the flow history between time 0 and the current time $t$. We will thus refer to this boundary condition as the "fully transient RCR boundary condition".

In this model, the operators defining the coupling terms on $\Gamma_B$ in equation (5) are:

$$\int_{\Gamma_B} \hat{\vec{w}} \cdot \underline{M}_m(\hat{\vec{v}}, \hat{p}) \cdot \hat{\vec{n}} ds = -\int_{\Gamma_B} \hat{\vec{w}} \cdot \hat{\vec{n}} \left( R \int_{\Gamma_B} \hat{\vec{v}} \cdot \hat{\vec{n}} ds + \int_0^t \frac{e^{-(t-t_1)/\tau}}{C} \int_{\Gamma_B} \hat{\vec{v}}(t_1) \cdot \hat{\vec{n}} ds dt_1 \right) ds$$

$$\int_{\Gamma_B} \hat{\vec{w}} \cdot \underline{H}_m \cdot \hat{\vec{n}} ds = -\int_{\Gamma_B} \hat{\vec{w}} \cdot \hat{\vec{n}} \left( \left( \hat{P}(0) - R \int_{\Gamma_B} \hat{\vec{v}}(0) \cdot \hat{n} d\Gamma - \hat{P}_d(0) \right) e^{-t/\tau} + \hat{P}_d(t) \right) ds \tag{7}$$

The continuous set of equations given by (5) is now fully characterized. These equations are discretized in space using a stabilized finite element method described in (Vignon-Clementel et al. 2006) and in time using a semi-implicit generalized α-method adapted for fluid-solid interaction (see (Figueroa et al. 2006)).

## 3    Numerical simulations and results

In this section, we first present results corresponding to the verification and initialization of the model. Then, we apply the model to different cases of normal or pathophysiological non-periodic phenomena. For each problem, we compute velocity and pressure in the domain of interest. A velocity profile is prescribed at the inlet surfaces of the various models while the downstream domains are represented with the fully transient RCR boundary condition described in the methods section, so that non-periodic phenomena can be captured. The distal pressure $P_d(t)$ is zero except for the patient-specific Glenn simulation of section 3.4. Simulations were run on an SGI Altix machine, using mesh adaptivity (Müller et al. 2005, Sahni et al. 2006), a typical time step of 0.4 – 0.8 ms, and 2-8 nonlinear iterations per time step. The time step size is set by the deformability requirements, except for the stenosis case where the turbulent jet necessitates a small time step. A Newtonian approximation is assumed with a viscosity of 0.04 g/(cm·s). The density of blood is 1.06 g/cm³. Poisson's ratio is 0.5, the wall density 1.0 g/cm³, the shear correction parameter is 5/6. The values of the material parameters are all physiologically realistic.

### 3.1    Verification and initialization

In this problem, we consider a pulsatile simulation in a simple deformable carotid artery model (straight cylindrical tube with cross-sectional area ($0.28$ cm²), wall thickness ($0.03$ cm) and length ($12.6$ cm)) where the input flow (mean flow of 6.5e⁶ cm³/s) is periodic and contains only ten frequencies. The Young's modulus of the vessel wall is $4.07e^6$ dyn/cm². This value is such that a maximum deformation of 5% is obtained with a physiologic range of pressures. The outlet boundary condition values used for this simulation were 2833.9 dyn·s/cm⁵ for $R$, 17678.8 dyn·s/cm⁵ for $R_d$ and $6.35e^{-6}$ cm⁵/dyn for $C$, to get a physiological range of pressure. The solutions were obtained using a 45,849 linear tetrahedral element and 9,878 node mesh with a time step of 0.8 ms. The simulations were run for a number of cardiac cycles, until a periodic solution was reached.

We have verified that the RCR relationship imposed as a boundary condition was numerically satisfied and checked for mass conservation between the inlet and the outlet. We extracted the flow rate (integration of the velocity times the normal over the outlet surface) and the mean pressure (integration over the outlet surface of the pressure, divided by its surface area) for the last cycle of the simulation. We then computed their Fourier transforms $Q(\omega)$ and $P(\omega)$ and analyzed the frequency content of their modulus and phase (Figure 2). Figure 3 shows the excellent agreement between the impedance derived from the numerical solution ($P(\omega)$ divided by $Q(\omega)$) and the theoretical prescribed RCR impedance. Note that only the first ten frequencies are relevant in this problem, since there is virtually no higher frequency information in the solution. We have also verified that periodicity was achieved as expected. Figure 4 illustrates the convergence to a periodic solution of the pressure waveform when the simulation is started from an initial pressure significantly lower than that of the periodic solution. While flow at the outlet typically reaches a periodic solution in two or three cycles, pressure takes longer to converge to a periodic solution. This is due to the prescription of flow at the inlet and a relationship between flow and pressure at the outlet (but not the pressure itself) that includes the memory of the flow rate over time (with the time constant $\tau$ =1.1 s).

Note that the number of cycles needed for convergence of the solution is highly sensitive to the initial pressure value and the type of boundary condition. This is especially true for this RCR boundary condition that takes into account the history of the flow rate for the entire time. It is possible to start from an initial value that is close to the periodic solution if the transient solution is started at a point in the cycle with a flow that is very close to the mean flow and an initial outlet pressure value is set based on the pressure field from a steady, rigid or deformable simulation corresponding to the mean flow. For example using an initial pressure of 97 mmHg instead of 68 mmHg would make the simulation begin with cycle 4 of Figure 4. Such a choice of starting flow and pressure has been made for all subsequent simulations.

### 3.2    *A simple case of non periodic input flow: an idealized carotid model with heart rate variability*

Flow and pressure in the carotid artery are naturally aperiodic due to heart-rate variability, breathing and other physiological factors (Holdsworth et al. 1999, Laleg et al. 2007). In this section, we demonstrate the ability of the methods described above to model flow *and* pressure aperiodicity due to heart rate variability.

#### 3.2.1    *Input data and material parameters*

The cross-sectional area (0.24 cm$^2$), wall thickness (0.09 cm) and length (3.5 cm) of the straight portion of the left common carotid of a young healthy female subject (27 years old) were measured with pulsed-Doppler ultrasound (Philips IU-22) and averaged over time. An idealized geometric model (straight cylindrical vessel) was then constructed based on these measurements.

The mean velocity (i.e. averaged across the Doppler beam) at the inlet of the common carotid artery was also recorded over time, showing a significant cycle-to-cycle variability (see the red tracing of Figure 5). The flow rate at the inlet face of the model was obtained by multiplying the instantaneous mean velocity by the average cross sectional area of the carotid at this location. This approximation is justified by the fact that the displacement of the wall was small and that the velocity profile was essentially axisymmetric. The measured cycle-averaged peak velocity was within the standard deviation of the value reported by Holdsworth et al. (Holdsworth et al. 1999). The calculated flow rate (7.4 cm$^3$/s) was higher than the average value reported by Holdsworth et al. (6.0 cm$^3$/s) and Marshall et al. (6.1 cm$^3$/s) (Marshall et al. 2004). The instantaneous flow rate was then mapped into a parabolic velocity profile (justified by the fact that the Womersley number was ~ 2). This time- varying parabolic velocity profile was the inlet boundary condition considered for the numerical simulation. The simulation was run for the seven complete cardiac cycles shown in Figure 5.

The Young's modulus for the vessel wall ($4.18e^6$ dynes/cm$^2$) was chosen so that the numerical radial deformation matched the measured radial deformation (4%) of the wall, averaged over eight cardiac cycles, for the arm cuff measured pressure pulse (40 mmHg).

For the outlet boundary, we used the fully transient RCR boundary condition described in the methods section. The total resistance was calculated as the ratio of the time-averaged measured pressure (defined as 1/3 times the systolic pressure + 2/3 times the diastolic pressure) divided by the average input flow rate. The proximal resistance was calculated using Westerhof's data (Westerhof et al. 1969) for the segments downstream of the common carotid artery, considering that this computational model accounts for half of the first segment of the common carotid artery. The capacitance value was then adjusted so that the computed pulse pressure reasonably matched the measured pulse pressure (40mmHg). Hence the values used for this simulation were 1117.1 dyn·s/cm$^5$ for $R$ , 12144.1 dyn·s/cm$^5$ for $R_d$ and $3.18e^{-5}$ cm$^5$/dyn for $C$ .

Finally, we considered a computational mesh of 191,300 linear tetrahedral elements. The time step was 0.4 ms, dictated by the wall deformation requirements.

### 3.2.2    Numerical results

Figure 6 shows the flow and pressure values at the outlet face of the model. Since the prescribed input flow rate is aperiodic, the resulting pressure is also aperiodic, following the heart rate variability. Furthermore, and as expected, pressure waves lag flow waves. The pulse pressure, pressure levels  and characteristic shape of the obtained carotid pressure wave favorably correspond to physiological values see in the carotid artery of normal subjects reported by different studies in (Nichols and O'Rourke 2005): the time to the first systolic peak is around 0.13 s , there is the characteristic carotid high second pressure peak and the augmentation index is around -20% (Nichols and O'Rourke 2005). This example clearly demonstrates the possibility of modeling flow and pressure in the context of heart rate variability, enabled by this boundary condition that does not assume periodicity of the solution.

In order to study the dependence of the solution on the initial condition, the simulation was run a second time starting from the second cycle measured with the ultrasound data. This question is particularly relevant since the input flow rate, and thus the results, are aperiodic in time. As shown in Figure 7, the pressure waveforms at the outlet converge to the same solution after two cycles, thus demonstrating the relatively small impact of the initial condition on the long-term behavior of the solution.

### 3.3    Non periodic input flow in a carotid bifurcation with a stenosis in the internal carotid: a patient-specific complex geometry with transitional flow

This case presents a more complex geometry: the stenosis in the internal branch of this patient-specific carotid bifurcation induces intricate flow patterns as flow rate increases during systole and rapidly decelerates afterwards.

In this example, there is a 68% diameter-reduction stenosis which is considered to be in the lower range of a ''critical stenosis'' (75% and above diameter reduction). This level of restriction is usually symptomatic and in many cases may require an intervention to alleviate the pressure drop and the lower flow induced by the stenosis. The same typical inlet flow was imposed as in the previous example, scaled down to 5 cm$^3$/s, to take into account the chronic reduction of flow resulting from adaptation to the presence of the stenosis. As suggested by the previous example, the first cardiac cycle was imposed three times to achieve a periodic solution before beginning the simulation of the non-periodic part representing seven cardiac cycles.

Only the non-periodic part is shown in the results. The RCR boundary conditions were based on the one in 3.2: the total resistance was lowered in conjunction with the flow which was distributed according to the ratio 30% for the external carotid and 70% for the internal carotid (Nichols and O'Rourke 2005). Figure 8 summarizes these parameters. Aperiodicity here is a consequence of both the inlet flow and the stenosis. To study the interaction of the non-periodic inflow and the stenosis without the additional interaction with the vessel wall response, the wall was considered as rigid. From a numerical point of view, we utilized a finite element mesh consisting of 330,710 linear tetrahedral elements and a time step for the pulsatile simulation of 0.5 ms.

The constriction generates a downstream transitional and non-axisymmetric flow (see Figure 9). Flow and mean pressure at the outlets vary from one cycle to the next as shown in Figure 10. Mean pressure at the outlet is 70 mmHg in the internal carotid and 90 mmHg in the external carotid, underlying the strong effect of the stenosis. Note that this effect varies over the cardiac cycle and from cycle to cycle: the pressure and the flow differences between the two outlets (external – internal) vary respectively from 0 to 66 mmHg and -5.5 $cm^3$/s to 2.3 $cm^3$/s. The mean outlet flows are for the internal and external carotids 64% and 36% of the inlet flow respectively. This distribution is different from the ratio (70/30) imposed through the outlet boundary condition, stressing the non-negligible effect of the stenosis. The pressure difference between the two outlets is almost zero at end diastole, and varies in systole considerably from one cycle to the next (Figure 10). Note also the flow variability in the two branches between different cycles. This geometry is such that the maximum peak systole flow difference between two different cycles (10% – see Figure 8) increases in the external carotid (this value is 40% for the external carotid flow – see Figure 10). This suggests that the stenosis amplifies the variability due to the inflow.

To study the non-periodicity due to geometry in absence of other variability, another simulation was run with a periodic inflow (imposing the first cardiac cycle of the previous signal repeated for seven cycles). As shown in Figure 11, a negligible non-periodicity was found between different cycles in the mean outlet flow and pressure. We therefore conclude that for this level of stenosis and flow, most of the observed outflow non-periodicity is due to the non-periodicities in the inlet flow.

### 3.4 *Non periodic input flow in a complex multi-branched geometry: a patient-specific Glenn model with flow variations due to respiration and cardiac rhythm*

Congenital heart defects are the most common cause of death from birth defects. "Single ventricle physiology" (e.g. hypoplastic left heart or tricuspid atresia) is one of the most complex and least understood forms of congenital heart disease. Treatments for these conditions require a multi-staged surgical approach. The second (Glenn) and third (Fontan, total cavopulmonary connection - TCPC) stages of these procedures involve connecting the venae cavae (Glenn – superior vena cava; Fontan - inferior vena cava) to the right pulmonary artery, thus bypassing the heart. As flow to the pulmonary arteries is then passive in nature, optimal architecture and physiology are paramount for success. The resulting geometry for the Glenn (Figure 12), and for the Fontan are complex, creating unsteady flow structures even when steady inlet flow is prescribed in in-vitro experiments or in numerical simulations modeling this physiology (DeGroff et al. 2005, Pekkan et al. 2005). In a Glenn procedure, flow from the superior vena-cava (SVC) is directed almost perpendicularly to the pulmonary arteries and gets distributed into the two pulmonary trees, sometimes creating very complex flow and unique hemodynamic conditions that are still poorly understood. Furthermore, cardiac and respiration effects are asynchronous in general. Thus, as shown in Figure 13, blood flow is noticeably aperiodic in time.

Although progress has been made in modeling such systems (see for example (Guadagni et al. 2001, Migliavacca et al. 2003, Pekkan et al. 2005)), the combined effects of respiration and cardiac rhythm, although significant, have rarely been taken into account (Marsden et al. 2007) and, to our knowledge, the resulting aperiodicity has not been modeled. Here, special attention has been given to the boundary conditions in order to model this effect.

The patient-specific data presented for this example come from clinical data of a 3 year old child, shortly before undergoing the Fontan procedure. In Figure 12, we show the geometrical model constructed from magnetic resonance angiography (MRA) data for the larger arteries and cineangiography to add to the model the outlets that were difficult to see on the MRA. The SVC inlet flow rate was constructed from the ultrasound time-varying velocity (Figure 13) scaled by the SVC flow from the catheterization report, assuming a parabolic profile. The mean flow rate is 21.5 cm$^3$/s. The average (estimated from the ultrasound data) heart rate and respiratory rate were 115 bpm and 40 bpm, respectively. The simulation was initialized by imposing at the inflow the first cardiac cycle two times and then run with the eight cardiac cycles of Figure 13. Fully transient RCR outlet boundary conditions were prescribed to accommodate the aperiodicity of the flow in the system. The different parameters for the twelve outlets were chosen using a tuning algorithm (Spilker et al. 2007) that generates an impedance for each outlet based on the targeted measured mean pressure in the SVC, the distribution of the pulmonary segments among the different outlets and morphometric data for pulmonary vascular trees. Table 1 gives the RCR values obtained for each outlet. Downstream pressure was taken as the mean value of the left atrial pressure measured in the catheterization laboratory (5.5 mmHg). Constant wall thickness (0.035 cm) was calculated as described in (Podesser et al. 1998) based on the inlet radius. The Young's modulus for the vessel wall (9.4e$^5$ dyn/cm$^2$) was chosen so that the numerical displacement matched the averaged radial deformation of the SVC wall measured by MRI (2.5%), for the measured pressure pulse of a representative cardiac cycle (1 mmHg). The computational mesh has 820,000 linear tetrahedral elements and 160,000 nodes. The time step adopted is 0.8 ms.

Figure 14 shows a snapshot of pressure, streamlines of velocity and wall shear stress (WSS) during diastole. Intricate flow patterns are observed in the right pulmonary artery, due to the complexity of the geometry. Similar complex flow features were observed in the catheterization laboratory during angiography. The wall shear stress map shows physiologic levels and realistic pressure levels and pressure gradients are also achieved. The calculated mean pressure is 10.7 mmHg, and the averaged gradients from SVC to LPA and SVC to RPA are 1.4 mmHg and 0.1 mmHg, respectively. These predicted values compare extremely well with the values reported in the catheterization report (10 mmHg, 1-2 mmHg and 0 mmHg respectively). Furthermore (see Figure 15), the computed pressure, averaged over the cross-sectional area of the SVC, is shown as a function of time, along with the prescribed inflow for approximately two respiratory cycles. Flow and pressure follow the cardiac and respiratory rhythms, but are not in complete synchrony due to the capacitance of the pulmonary trees. The amplitude of the pressure curve over two respiratory periods is in good agreement with the recording of the pressure catheter at this anatomic location in the catheterization laboratory (see Figure 16). The differences in duration are probably due to that when the child was in the catheterization laboratory his heart rate (HR 94 bpm) was lower than when the ultrasound data (on which the simulation was based) was obtained (HR 115bpm).

Table 1: outlet boundary condition values for each branch. The proximal and distal resistances are in dyn·s/cm$^5$ and the capacitances in cm$^5$/dyn. There are seven outlets for the right lobes and five outlets for the left lobes (the outlets are ordered in their order of branching off the right and left pulmonary arteries, from top to bottom, consistently with Figure 14).

| Outlet | RUL | RSL | RBL | RML | RLL_2 | RLL | RPA | LUL | LSL | LML | LLL_2 | LPA |
|---|---|---|---|---|---|---|---|---|---|---|---|---|
| R/10$^2$ | 3.03 | 9.08 | 4.54 | 9.08 | 9.08 | 9.08 | 9.08 | 1.59 | 7.97 | 3.98 | 7.97 | 7.97 |
| C*10$^5$ | 7.69 | 2.56 | 5.13 | 2.56 | 2.56 | 2.56 | 2.56 | 15.56 | 3.11 | 6.22 | 3.11 | 3.11 |
| Rd/10$^3$ | 1.72 | 5.17 | 2.59 | 5.17 | 5.17 | 5.17 | 5.17 | 0.77 | 3.84 | 1.92 | 3.84 | 3.84 |

## 4 Discussion

An important aspect of the implemented boundary condition lies in its implicit nature whereby neither pressure nor flow are directly prescribed. Instead, a relationship between the two is enforced and therefore, this relationship can be employed for short-term predictions of surgeries. Furthermore, in the examples presented above, the results show good agreement with targeted values and the available experimental data. Therefore, the limitations presented in the following paragraphs do not hinder the main purpose of modeling non-periodic phenomena, especially considering that most of these phenomena are also present when modeling periodic states. Rather, the discussion of the next paragraph should be seen as suggestive of future research.

The first limitation of the methodology described here results from the difficulty of simultaneously measuring in *vivo* pressures and flow rates. For example, for the Glenn patient studied here, input flow was measured with echocardiography whereas pressure was measured in the catheterization laboratory when the child was under anesthesia and respiration control. This can explain the discrepancies seen in the pulsatility of the computed and measured pressure in the SVC. Furthermore, the relationship between pressure and flow imposed as a boundary condition is a simple model (here the RCR model) and is thus not determined from the patient's time-varying measurements. However, such measurements present some technical challenges and are feasible only in a very limited number of research cases. More complex zero-dimensional time-varying boundary conditions could be used to represent the downstream vasculature, although the patient-specific determination of their parameters is not obvious.

As mentioned in the introduction, a pure resistance boundary condition affects the pressure pulse and rate of decay (Vignon-Clementel et al. 2006, Vignon and Taylor 2004). Both the pressure pulse and decay are over-predicted. Consequently, the vessel wall displacements will be accordingly over-estimated. A standard RCR model can be used to relate pressure and flow at the outlet. This standard RCR model computes the impedance in the frequency domain from an analogy with electric circuits which assumes periodicity of the solution. It takes into account the history of the flow rate during one cardiac cycle only and thereby neglects the transient term in (6) – see e.g. (Vignon-Clementel 2006). In the rest of the text, it will be referred as the "periodic RCR boundary condition" by contrast to the "fully transient RCR boundary condition" defined in the Methods section. If the aperiodicity is not too strong, as in section 3.2, then the solutions for the periodic RCR boundary condition and the fully transient boundary condition are not significantly different (see the outlet pressure wave forms in the last three cardiac cycles of the simulation shown in Figure 17). Moreover, the computing time is also not significantly different.

By contrast, when aperiodicity is larger, there can be more significant differences between the two types of boundary conditions. We consider here the example of the carotid bifurcation with a stenosis (see section 3.3), imposing a non-periodic inflow. The numerical solutions for the periodic RCR boundary condition and fully transient RCR boundary condition are significantly different for the cycles where aperiodicity is the most pronounced. Figure 18 shows the flow and pressure wave forms at the outlet for the two types of boundary conditions. The flow wave forms are almost identical but the pressure wave forms exhibit significant differences, especially in the internal carotid: there is 10% maximum pressure difference between the periodic and fully transient RCR boundary conditions. This difference is of the same order of magnitude as the cycle to cycle inflow variation: as can be seen in Figure 8, the maximum peak systole flow difference between two different cycles is 10%. Moreover, the plots in Figure 18 show that pressure is overpredicted for some cycles and underpredicted for others when the periodic RCR boundary condition is used. Figure 19 shows the velocity magnitude and pressure differences at time 3.6 s in the computational domain: the velocity patterns, especially downstream of the stenosis are different and the pressure maps exhibit remarkable differences.

From a numerical point of view, for all the simulations of this section, the effort to solve the non-linear equations on each step to the same tolerance was somewhat lower for fully transient RCR relative to periodic RCR simulations. Moreover, based on the conclusion of case 3.2, an initialization was done for every simulation by imposing at the inflow a first cardiac cycle repeatedly three times to achieve periodicity of the solution prior to simulate the rest of the desired periodic or non-periodic inflow. Such an initialization exhibited a larger magnitude of the pressure variation and a larger number of cycles for the periodic RCR than for the fully transient RCR simulations. By contrast, when the solution is periodic, there were no differences (both less than 0.1%) in the residual or number of iterations per time step. The fully transient RCR boundary condition appears thus as numerically more favorable than the periodic RCR boundary condition in the presence of aperiodic phenomena.

In addition, there is the issue of numerical periodicity. In the first numerical example it was verified that periodicity was achieved when expected (see Figure 3). Note that "when expected" refers to the fact that if a periodic input flow is prescribed, in general flow and pressure are periodic in time in the modeled domain. In certain geometries however (DeGroff et al. 2005, Nichols and O'Rourke 2005, Otsuka et al. 1997, Pekkan et al. 2005, Sherwin and Blackburn 2005), significant unsteady flow can occur even if the inlet flow is steady. This can be due to the complex geometry of the model that induces intricate flow patterns as flow rate decelerates after peak systole (transition to turbulence in a post-stenotic region) or due to complex flow forming when two (even steady) input jets collide into a complex geometry such as the Fontan circulation (Vignon-Clementel 2006).

To sum up, in the cases where non-periodicity is important, we postulate that the use of a boundary condition that makes no assumption on periodicity of the flow or pressure is a more sensitive choice.

The formulation considered here has been verified using Womersley's deformable wall solution with an outflow impedance boundary condition that assumes periodicity of flow and pressure (Figueroa 2006). Verification of the fully transient RCR boundary condition and initialization approach has been done in sections 3.1 and 3.2. However, validation should be done using *in vitro* and *in vivo* data (Ku et al. 2000, Ku et al. 2001, Ku et al. 2002). We have begun to compare the results with clinical pressure measurements (Marsden et al. 2007) but this matter is subject of current and future work. In particular a flow-phantom in vitro-experiment could be used to validate the flow and pressure obtained with a variable pulsatile inflow and fully transient RCR boundary conditions.

## 5    Conclusions

We first verified the numerical model in a simple geometry. In the example of an idealized carotid geometry with aperiodic input flow, heart rate variability was studied. We demonstrated that flow *and* pressure variations due to naturally varying heart rate can be simulated and discussed the sensitivity to the initial conditions. In the example of the stenotic carotid bifurcation, we simulated flow and pressure variations due to naturally varying heart rate in the presence of a stenosis. In case of time varying inflow, the influence of the stenosis was studied: it created large flow and pressure variations in the two branches, suggesting an amplification of the natural time variability of the inflow. We showed that when the inflow is periodic, there is for these parameters very little aperiodicity in the solution. In the example of an aperiodic input flow in the complex multi-branched geometry of a patient-specific Glenn model, we demonstrated that flow and pressure variations due to naturally varying respiration and cardiac rhythms can be simulated. This implies that if time-varying patient data is available, there is no need to construct a model to synchronize respiratory and cardiac rhythms as we previously described (Marsden et al. 2007). In the discussion section, we showed that boundary conditions that make no assumption of periodicity of pressure and flow exhibit different numerical solutions than those assuming periodicity of the solution. We thus postulate that they are more appropriate boundary conditions, especially if the aperiodicity is large.

Despite the aforementioned limitations, the numerical simulations presented in this paper illustrate the ability of the described methodology to model important transient phenomena such as heart-rate variability, respiratory and cardiac asynchronous variations, and non-periodic phenomena due to complex geometries and flows. The present approach can be used to incorporate sophisticated lumped parameter models describing time-varying phenomena such as cardiac perfusion (Spaan et al. 2000), autonomic regulatory mechanism (Ursino and Magosso 2003), the baroreflex control linked to blood pressure and heart rate variability (Olufsen et al. 2002, Olufsen et al. 2005, Olufsen et al. 2006, TenVoorde and Kingma 2000), effects of hypoxia (Ursino and Magosso 2001), respiration, exercise and their combinations (Magosso and Ursino 2002, Magosso et al. 2005). Moreover, this approach can be applied to other flow problems where pressure is particularly important, such as respiratory airflow in the lungs which is strongly coupled to the parenchyma and the diaphragm movement (Grandmont et al. 2005).

## 6    Acknowledgments

This material is based upon work supported by the National Science Foundation under Grant No. 0205741 and a predoctoral fellowship from the American Heart Association to Irene Vignon-Clementel. The authors gratefully acknowledge Dr. Farzin Shakib for the use of his linear algebra package AcuSolve™ (http://www.acusim.com) and the support of Simmetrix, Inc for the use of the MeshSim™ automatic mesh generator (http://www.simmetrix.com). The authors gratefully acknowledge the assistance of Bonnie Johnson for the acquisition of the carotid flow data, Dr. Frandics Chan for acquiring the subject-specific congenital MRI data, Adam Bernstein for constructing the congenital geometrical model, Ryan Spilker for the boundary condition tuning algorithm for that same example and Dr. Jeffrey Feinstein for his guidance on the Glenn pathophysiology.

## 7    References

Bertram CD. 1995. The dynamics of collapsible tubes. Symp Soc Exp Biol. 49:253-264.

Blanco PJ, Feijoo RA and Urquiza SA. 2007. A unified variational approach for coupling 3D-1D models and its blood flow applications. Computer Methods in Applied Mechanics and Engineering. 196(41-44):4391.

Bove EL, de Leval MR, Migliavacca F, Guadagni G and Dubini G. 2003. Computational fluid dynamics in the evaluation of hemodynamic performance of cavopulmonary connections after the Norwood procedure for hypoplastic left heart syndrome. Journal of Thoracic and Cardiovascular Surgery. 126(4):1040.

Brown DJ. 1996. Input impedance and reflection coefficient in fractal-like models of asymmetrically branching compliant tubes. IEEE Trans Biomed Eng. 43(7):715-722.

DeGroff C, Birnbaum B, Shandas R, Orlando W and Hertzberg J. 2005. Computational simulations of the total cavo-pulmonary connection: Insights in optimizing numerical solutions. Medical Engineering and Physics. 27(2):135-146.

Figueroa CA. A couple-momentum method to model blood flow and vessel deformation in human arteries: Applications in disease research and simulation-based medical planning [PhD thesis]. Stanford: Stanford University.

Figueroa CA, Vignon-Clementel IE, Jansen KE, Hughes TJR and Taylor CA. 2006. A coupled momentum method for modeling blood flow in three-dimensional deformable arteries. Computer Methods in Applied Mechanics and Engineering. 195:5685-5706.

Formaggia L, Gerbeau JF, Nobile F and Quarteroni A. 2001. On the coupling of 3D and 1D Navier-Stokes equations for flow problems in compliant vessels. Computer Methods in Applied Mechanics and Engineering. 191:561-582.

Frank O. 1899. Die Grundform des arteriellen Pulses. Zeitung für Biologie. 37:483-586.

Grandmont C, Maday Y and Maury B. 2005. A multiscale/multimodel approach of the respiration tree. In: New trends in continuum mechanics Bucharest: Theta. p. 147-157.

Guadagni G, Bove EL, Migliavacca F and Dubini G. 2001. Effects of pulmonary afterload on the hemodynamics after the hemi-Fontan procedure. Medical Engineering & Physics. 23(5):293-298.

Holdsworth DW, Norley CJD, Frayne R, Steinman DA and Rutt BK. 1999. Characterization of common carotid artery blood-flow waveforms in normal subjects. Physiological Measurement. 20:219-240.

Jensen OE. 1992. Chaotic oscillations in a simple collapsible-tube model. Journal of Biomechanical Engineering. 114(1):55-59.

Ku JP, Draney M, Lee WA, Zarins CK and Taylor CA. 2000. In vivo validation of a cardiovascular surgical simulation system. Annals of Biomedical Engineering. 28(SUPPL. 1):S-62.

Ku JP, Draney MT, Lee WA, Arko F, Zarins CK and Taylor CA. 2001. Comparison of in vivo MRI flow measurements and predicted flow simulation results. ASME Bioeng Div Publ BED. 50:769-770.

Ku JP, Draney MT, Arko FR, Lee WA, Chan FP, Pelc NJ, Zarins CK and Taylor CA. 2002. In vivo validation of numerical prediction of blood flow in arterial bypass grafts. Annals of Biomedical Engineering. 30(6):743-752.

Lagana K, Dubini G, Migliavacca F, Pietrabissa R, Pennati G, Veneziani A and Quarteroni A. 2002. Multiscale modelling as a tool to prescribe realistic boundary conditions for the study of surgical procedures. Biorheology. 39(3-4):359-364.

Lagana K, Balossino R, Migliavacca F, Pennati G, Bove EL, de Leval MR and Dubini G. 2005. Multiscale modeling of the cardiovascular system: Application to the study of pulmonary and coronary perfusions in the univentricular circulation. Journal of Biomechanics. 38(5):1129-1141.

Laleg T-M, Médigue C, Cottin F and Sorine M. Arterial blood pressure analysis based on scattering transform ii. In: IEiMaB Societys editor. 29th Annual International Conference Sciences and Technologies for Health - EMBC 2007; Lyon, France.

Magosso E and Ursino M. 2002. Cardiovascular response to dynamic aerobic exercise: A mathematical model. Med Biol Eng Comput. 40(6):660-674.

Magosso E, Ursino M, Ursino M, Brebbia CA, Pontrelli G and Magosso E. 2005. A theoretical study of the transient and steady-state cardiorespiratory response to exercise. In: 6th international conference on modelling in medicine and biology, Bologna, Italy: Ashurt Lodge. p. 17-26.

Marsden AL, Vignon-Clementel IE, Chan FP, Feinstein JA and Taylor CA. 2007. Effects of exercise and respiration on the hemodynamic efficiency in CFD simulations of the total cavopulmonary connection. Annals of Biomedical Engineering. 35(2):250-263.

Marshall I, Papathanasopoulou P and Wartolowska K. 2004. Carotid flow rates and flow division at the bifurcation in healthy volunteers. Physiological Measurement(3):691.

Migliavacca F, Dubini G, Bove EL and de Leval MR. 2003. Computational fluid dynamics simulations in realistic 3-D geometries of the total cavopulmonary anastomosis: The influence of the inferior caval anastomosis. Journal of Biomechanical Engineering. 125(6):805-813.

Müller J, Sahni O, Li X, Jansen KE, Shephard MS and Taylor CA. 2005. Anisotropic adaptive finite element method for modelling blood flow. Computer Methods in Biomechanics & Biomedical Engineering. 8(5):295-305.

Nichols WW and O'Rourke MF. 2005. McDonald's blood flow in arteries: Theoretical, experimental and clinical principles. 5th ed. New York: Oxford University Press.

Olufsen MS. 1999. Structured tree outflow condition for blood flow in larger systemic arteries. American Journal of Physiology. 276:H257-268.

Olufsen MS, Peskin CS, Kim WY, Pedersen EM, Nadim A and Larsen J. 2000. Numerical simulation and experimental validation of blood flow in arteries with structured-tree outflow conditions. Annals of Biomedical Engineering. 28(11):1281-1299.

Olufsen MS, Nadim A and Lipsitz LA. 2002. Dynamics of cerebral blood flow regulation explained using a lumped parameter model. Am J Physiol Regul Integr Comp Physiol. 282(2):R611-622.

Olufsen MS, Ottesen JT, Tran HT, Ellwein LM, Lipsitz LA and Novak V. 2005. Blood pressure and blood flow variation during postural change from sitting to standing: Model development and validation. Journal of Applied Physiology. 99(4):1523-1537.

Olufsen MS, Tran HT, Ottesen JT, Lipsitz LA and Novak V. 2006. Modeling baroreflex regulation of heart rate during orthostatic stress. Am J Physiol Regul Integr Comp Physiol. 291(5):R1355-1368.

Otsuka K, Nishimun Y, Kubo Y, Cornelissen G and Halberg F. Chronomes (rhythms, chaos, and age trends) of human heart rate variability in both genders. IEEE Computers in Cardiology.

Pekkan K, De Zelicourt D, Ge L, Sotiropoulos F, Frakes D, Fogel MA and Yoganathan AP. 2005. Physics-driven CFD modeling of complex anatomical cardiovascular flows-a TCPC case study. Annals of Biomedical Engineering. 33(3):284-300.

Podesser BK, Neumann F, Neumann M, Schreiner W, Wollenek G and Mallinger R. 1998. Outer radius-wall thickness ratio, a postmortem quantitative histology in human coronary arteries. Acta Anat (Basel). 163(2):63-68.

Quarteroni A, Ragni S and Veneziani A. 2001. Coupling between lumped and distributed models for blood flow problems. Computing and Visualization in Science. 4(2):111-124.

Quarteroni A and Veneziani A. 2003. Analysis of a geometrical multiscale model based on the coupling of ODEs and PDEs for blood flow simulations. Multiscale Modeling & Simulation. 1(2):173-195.

Sahni O, Muller J, Jansen KE, Shephard MS and Taylor CA. 2006. Efficient anisotropic adaptive discretization of the cardiovascular system. Computer Methods in Applied Mechanics and Engineering. 195(41-43):5634-5655.

Sherwin SJ and Blackburn HM. 2005. Three-dimensional instabilities and transition of steady and pulsatile axisymmetric stenotic flows. Journal of Fluid Mechanics. 533:297-327.

Spaan JA, Cornelissen AJ, Chan C, Dankelman J and Yin FC. 2000. Dynamics of flow, resistance, and intramural vascular volume in canine coronary circulation. Am J Physiol Heart Circ Physiol. 278(2):H383-403.

Spilker R, Feinstein J, Parker D, Reddy V and Taylor C. 2007. Morphometry-based impedance boundary conditions for patient-specific modeling of blood flow in pulmonary arteries. Annals of Biomedical Engineering. 35(4):546.

Steele BN, Olufsen MS and Taylor CA. 2007. Fractal network model for simulating abdominal and lower extremity blood flow during resting and exercise conditions. Computer Methods in Biomechanics and Biomedical Engineering. 10(1):39-51.

Stergiopulos N, Tardy Y and Meister JJ. 1993. Nonlinear separation of forward and backward running waves in elastic conduits. Journal of Biomechanics. 26(2):201-209.

Taylor CA and Hughes TJR. A multiscale finite element method for blood flow in deformable vessels. Proceedings of the1998 World Congress of Biomechanics; Sapporo, Japan.

Taylor CA, Draney MT, Ku JP, Parker D, Steele BN, Wang K and Zarins CK. 1999. Predictive medicine: Computational techniques in therapeutic decision-making. Comput Aided Surg. 4(5):231-247.

TenVoorde BJ and Kingma R. 2000. A baroreflex model of short term blood pressure and heart rate variability. Stud Health Technol Inform. 71:179-200.

Torii R, Oshima M, Kobayashi T, Takagi K and Tezduyar TE. 2006. Computer modeling of cardiovascular fluid-structure interactions with the deforming-spatial-domain/stabilized space-time formulation. Computer Methods in Applied Mechanics and Engineering. 195(13-16):1885-1895.

Urquiza SA, Blanco PJ, Venere MJ and Feijoo RA. 2006. Multidimensional modelling for the carotid artery blood flow. Computer Methods in Applied Mechanics and Engineering. 195(33-36):4002-4017.

Ursino M and Magosso E. 2001. Role of tissue hypoxia in cerebrovascular regulation: A mathematical modeling study. Annals of Biomedical Engineering. 29(7):563-574.

Ursino M and Magosso E. 2003. Role of short-term cardiovascular regulation in heart period variability: A modeling study. Am J Physiol Heart Circ Physiol. 284(4):H1479-1493.

Vignon-Clementel IE. 2006. A coupled multidomain method for computational modeling of blood flow [PhD thesis]. Stanford: Stanford University.

Vignon-Clementel IE, Figueroa CA, Jansen KE and Taylor CA. 2006. Outflow boundary conditions for three-dimensional finite element modeling of blood flow and pressure in arteries. Computer Methods in Applied Mechanics and Engineering. 195:3776-3796.

Vignon I and Taylor CA. 2004. Outflow boundary conditions for one-dimensional finite element modeling of blood flow and pressure waves in arteries. Wave Motion. 39(4): 361-374.

Westerhof N, Bosman F, De Vries CJ and Noordergraaf A. 1969. Analog studies of the human systemic arterial tree. Journal of Biomechanics. 2(2):121-143.

## 8   Figure captions

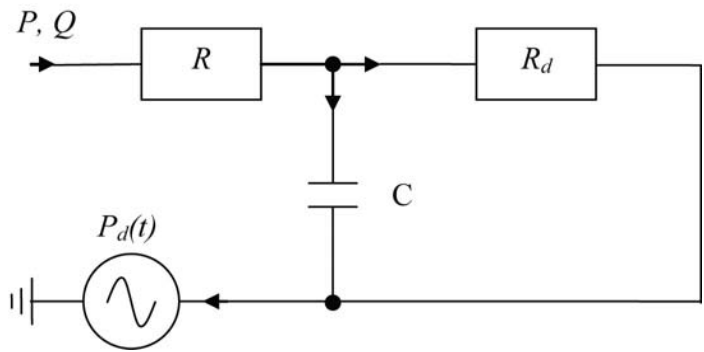

Figure 1: Windkessel electric analog.

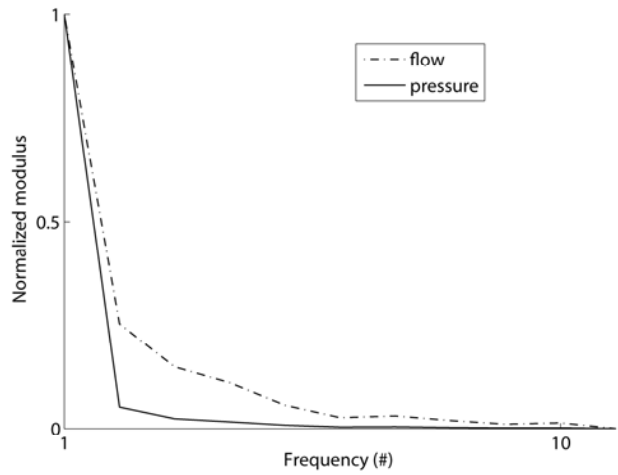

Figure 2: Frequency content of flow and pressure at the outlet of the deformable carotid model with a fully transient RCR boundary condition. Each modulus has been normalized by its zero Hz frequency value (first frequency on the graph).

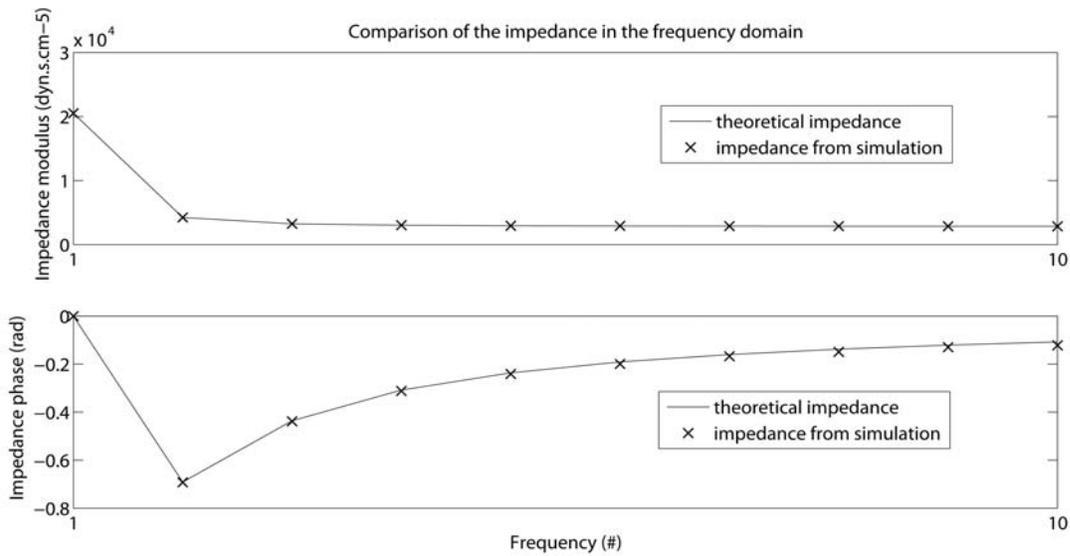

Figure 3: Impedance for the fully transient RCR boundary condition: comparison between theoretical and simulated values (P(ω) divided by Q(ω) for the last cycle of the simulation, ω being the frequency).

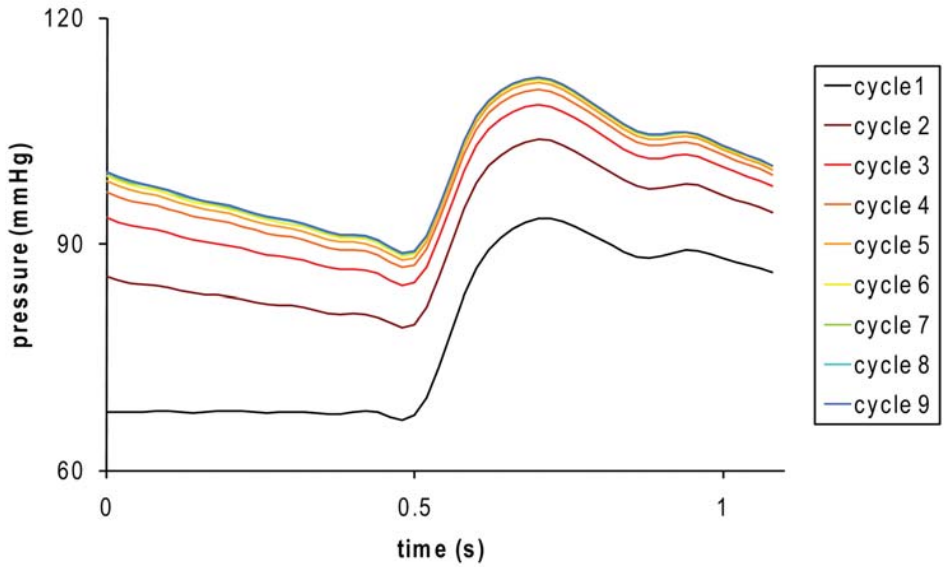

Figure 4: Convergence to periodicity of pressure for pulsatile flow in a deformable tube with prescribed inflow and fully transient RCR outlet boundary condition.

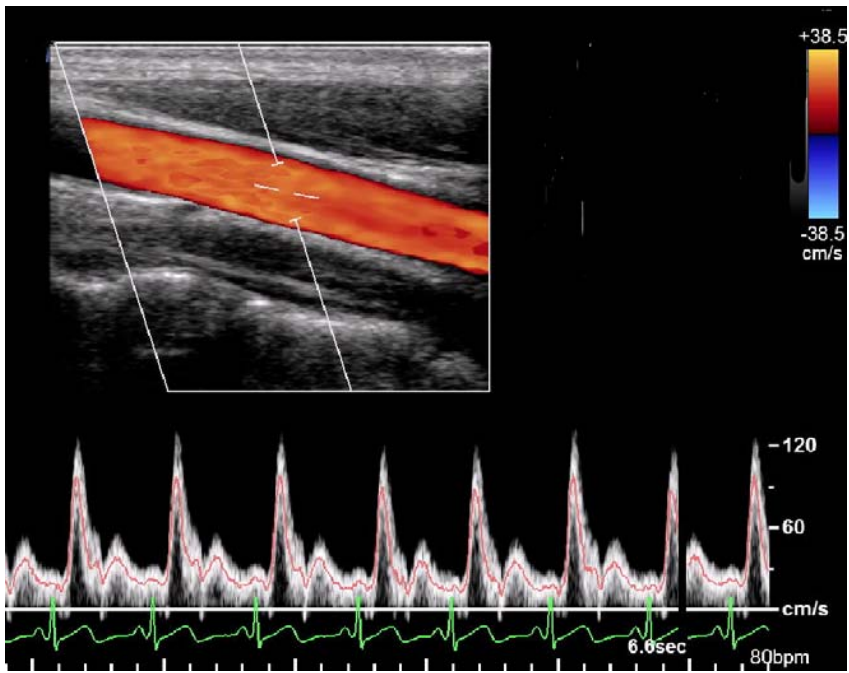

Figure 5: Ultrasound velocity measurement of a young normal female patient, showing the natural variability from one cardiac cycle to the next. The red curve indicates the mean velocity and the green curve the simultaneous ECG tracing that marks the timing of the different heart beats. The upper part shows the Doppler signal in a longitudinal cut of the carotid, indicating forward flow along and across the artery (see the color bar for the indication of the velocity value and direction).

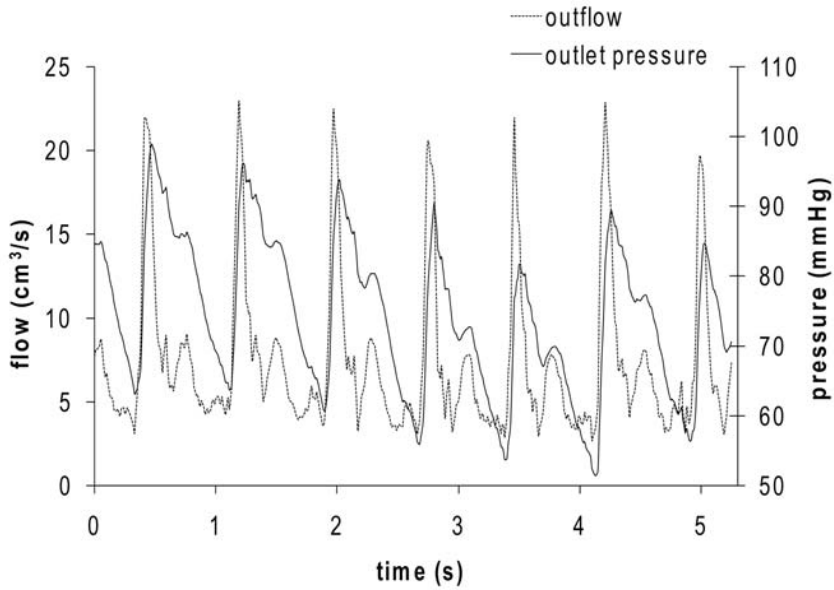

Figure 6: Non-periodic flow and pressure waveforms at the outlet of the carotid model, resulting from a pulsatile simulation over seven cardiac cycles with measured inlet flow and the fully transient RCR boundary condition.

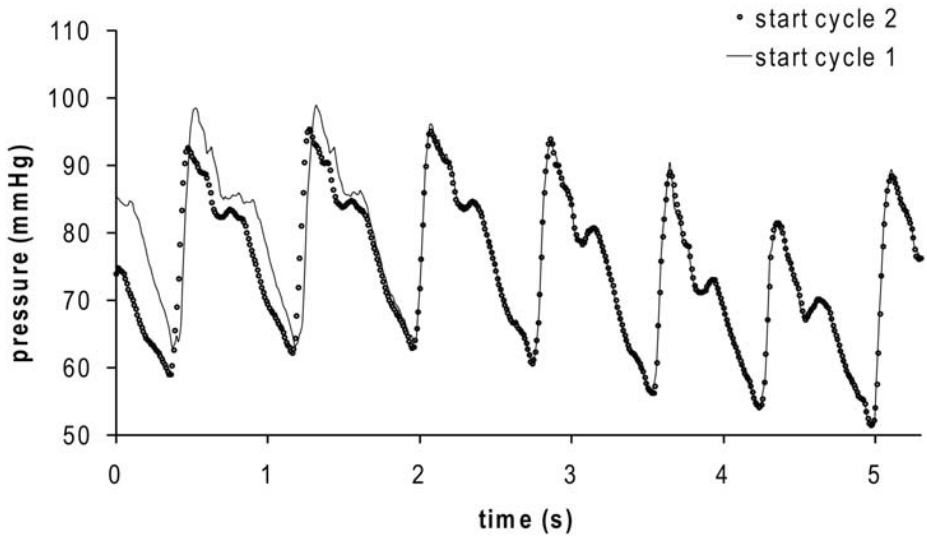

Figure 7: Comparison of the solution when starting from two different cycles, showing convergence to the same solution after a few cardiac cycles.

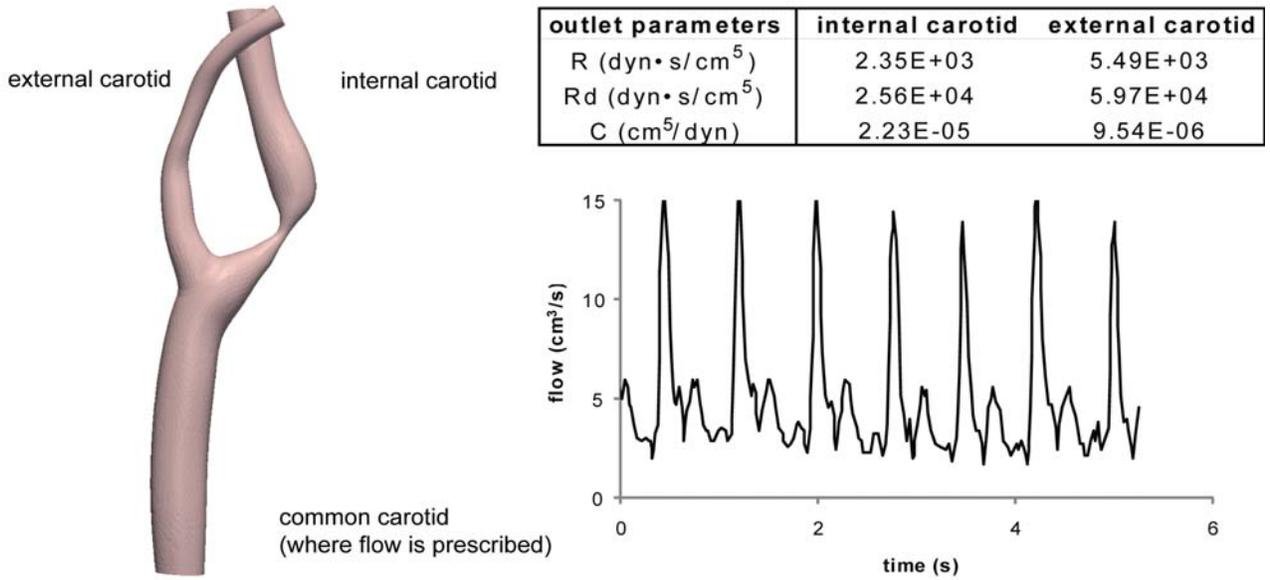

| outlet parameters | internal carotid | external carotid |
|---|---|---|
| R (dyn·s/cm$^5$) | 2.35E+03 | 5.49E+03 |
| Rd (dyn·s/cm$^5$) | 2.56E+04 | 5.97E+04 |
| C (cm$^5$/dyn) | 2.23E-05 | 9.54E-06 |

Figure 8: Model parameters for the carotid bifurcation. Patient-specific geometry on the left. Inlet flow wave form and RCR outlet parameters on the right.

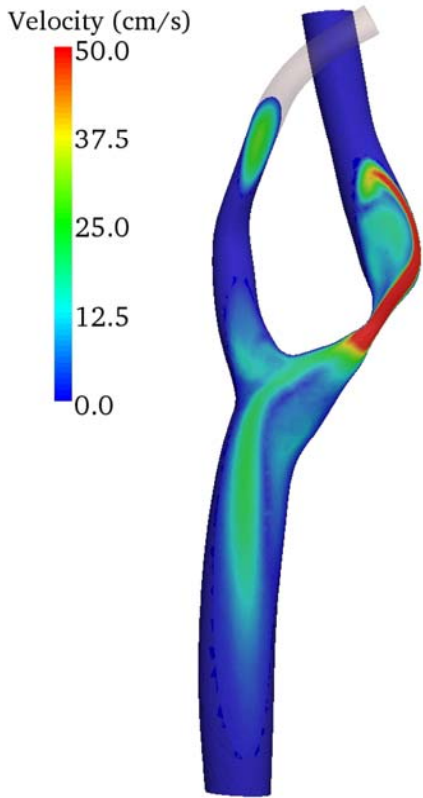

Figure 9: Velocity magnitude on a cut of the carotid bifurcation at time 3.6 s. Note the complex flow patterns, especially downstream of the stenosis.

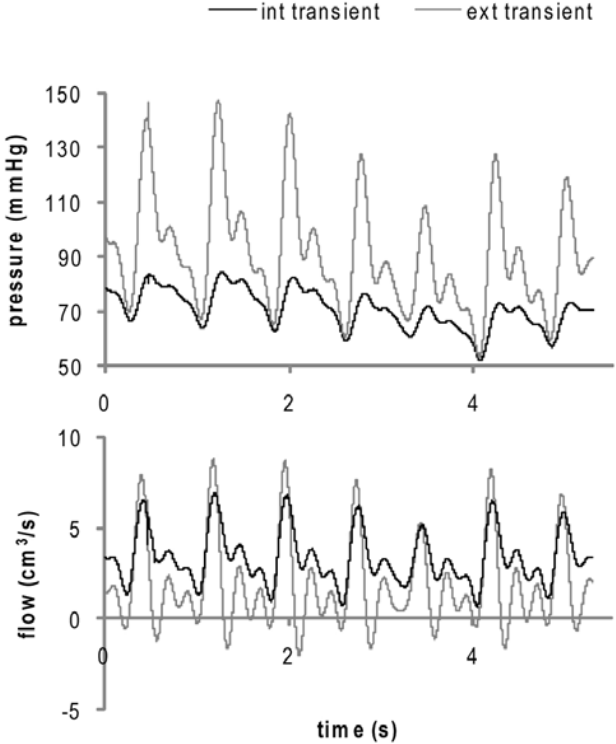

Figure 10: Mean pressure and flow over seven cardiac cycles at the outlet of the internal and external carotid artery branches.

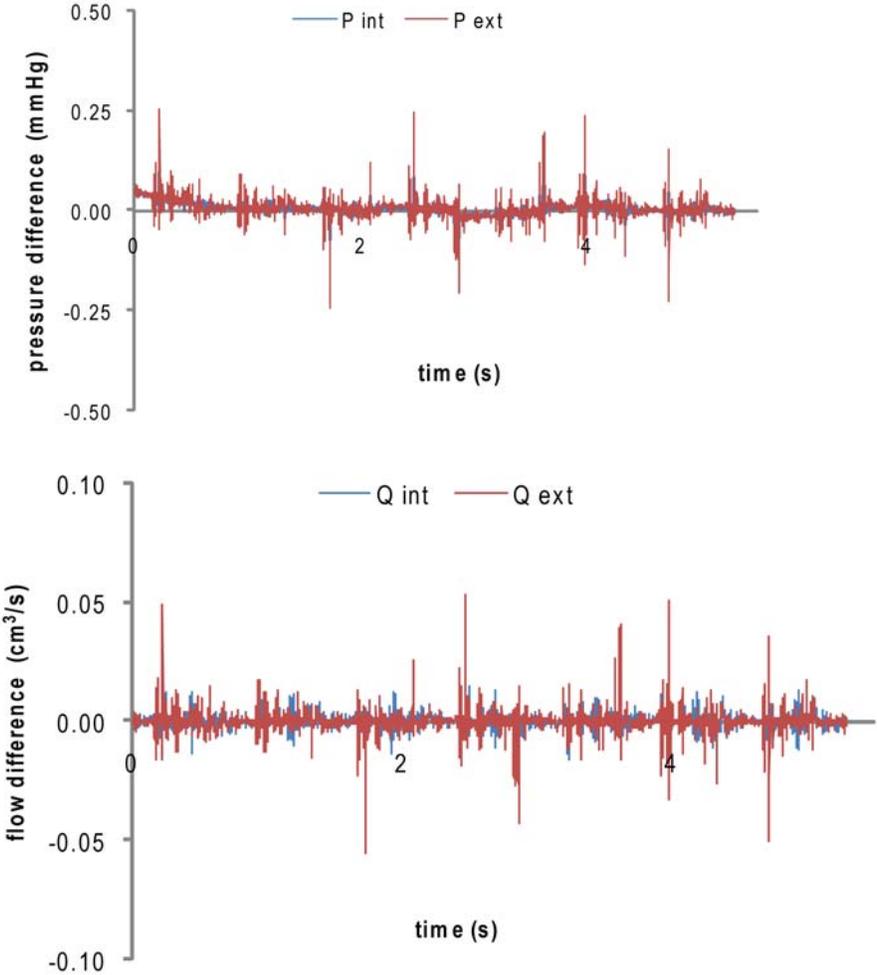

Figure 11: Mean pressure difference (P(t+T) – P(t)) and flow difference (Q(t+T)-Q(t)) as a function of time at the two outlets when a periodic inflow of period T is imposed. Negligible differences due to the non-periodicity generated by the stenosis can be seen.

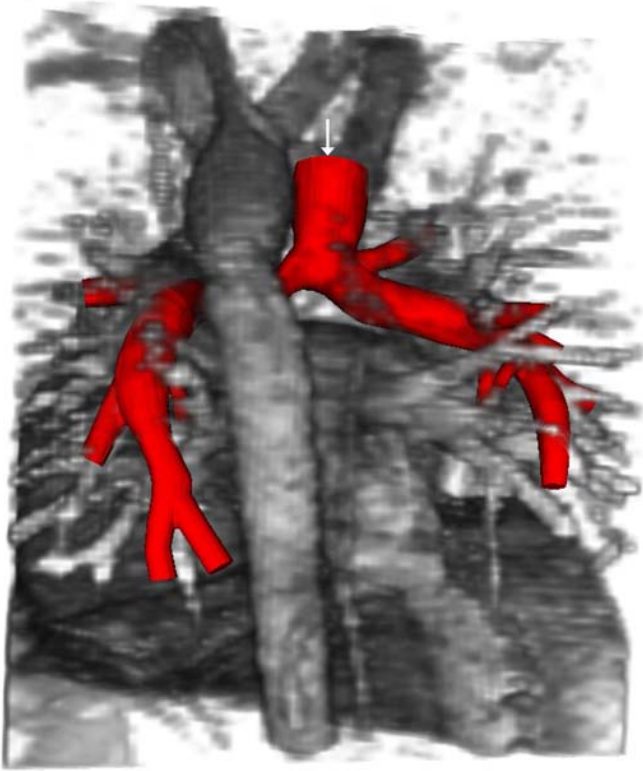

Figure 12: Geometric model (in red) reconstructed from and superposed on the magnetic resonance angiography (MRA) data of a Glenn patient. The arrow indicates the SVC inflow; all the other arteries are outlets.

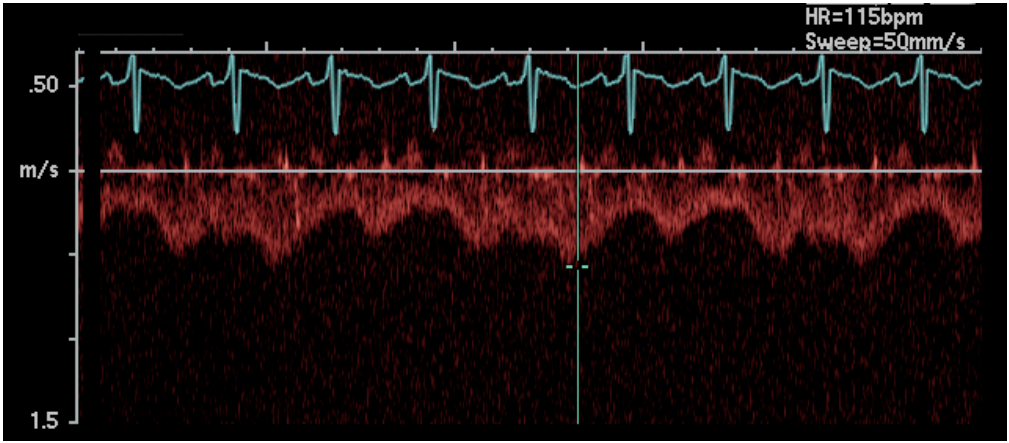

Figure 13: Velocity (in red) measured with echo-Doppler ultrasound in the SVC of a 3 year old Glenn patient showing cardiac and respiratory variations. The turquoise curve indicates the simultaneous ECG tracing that marks the timing of the different heart beats.

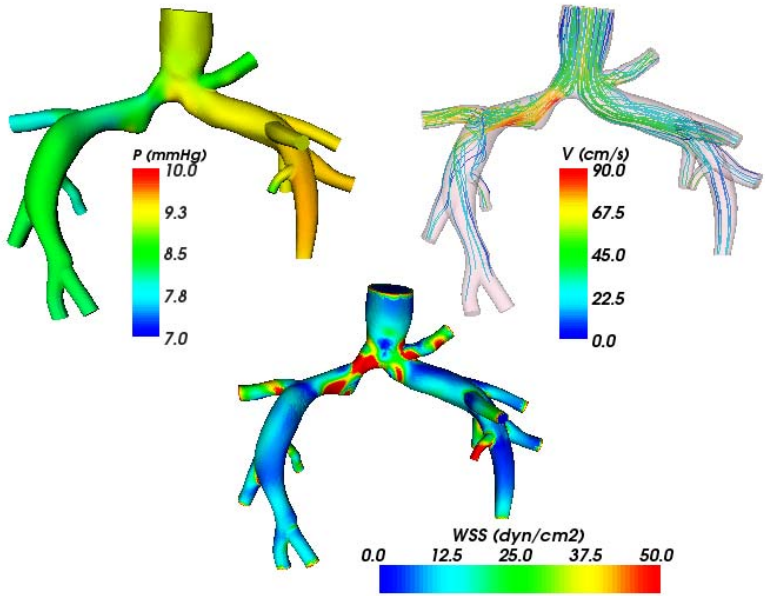

Figure 14: Pressure (P), streamline velocity magnitude (V) and wall shear stress (WSS) shown in diastole for the Glenn model. Realistic pressure magnitude and complex flow and wall shear stress patterns were obtained.

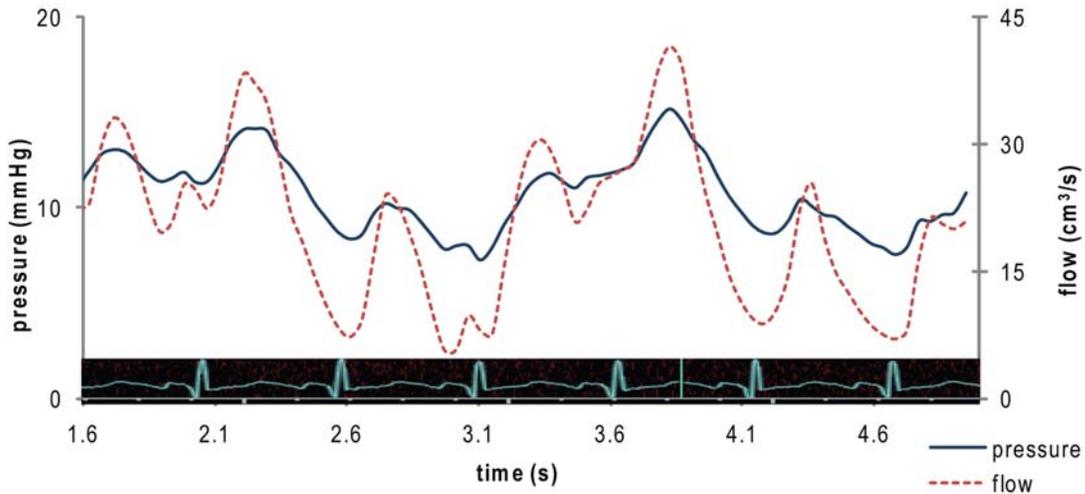

Figure 15: SVC pressure averaged over the cross-sectional area and prescribed inflow shown over approximately two respiratory cycles. Superimposed to the plots is the ECG tracing corresponding to the inflow measurement (Figure 13). The average (estimated from the ultrasound data) cardiac period and respiratory period were 0.5 s (time marks on the time axis) and 1.5 s respectively. Flow and pressure follow both the cardiac and respiratory rhythms. They are asynchronous due to the capacitance of the pulmonary vasculature.

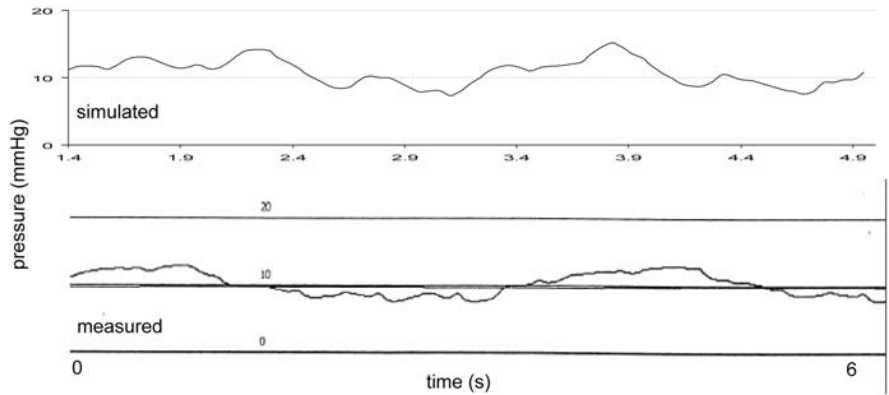

Figure 16: Comparison of simulated pressure over time with the measure by a pressure catheter in the SVC.

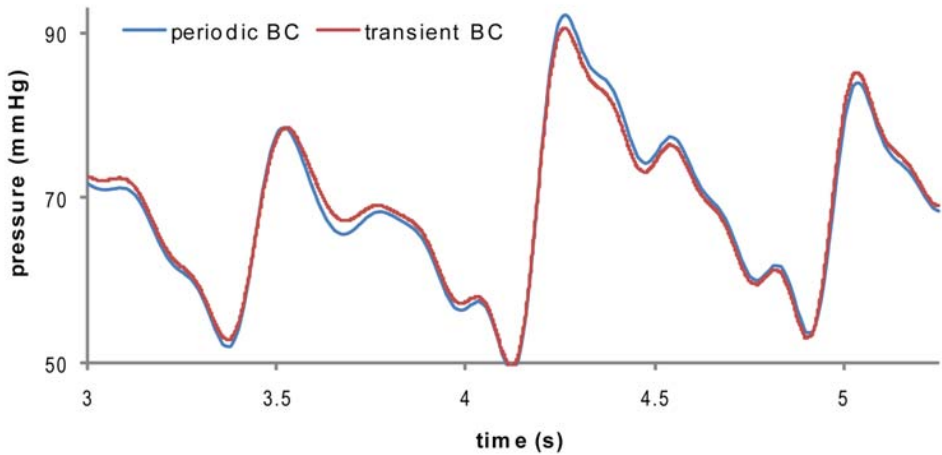

Figure 17: Comparison of pressure (averaged over the outlet boundary) for the periodic RCR boundary condition and the fully transient RCR boundary condition in the simple carotid example (as in section 3.2).

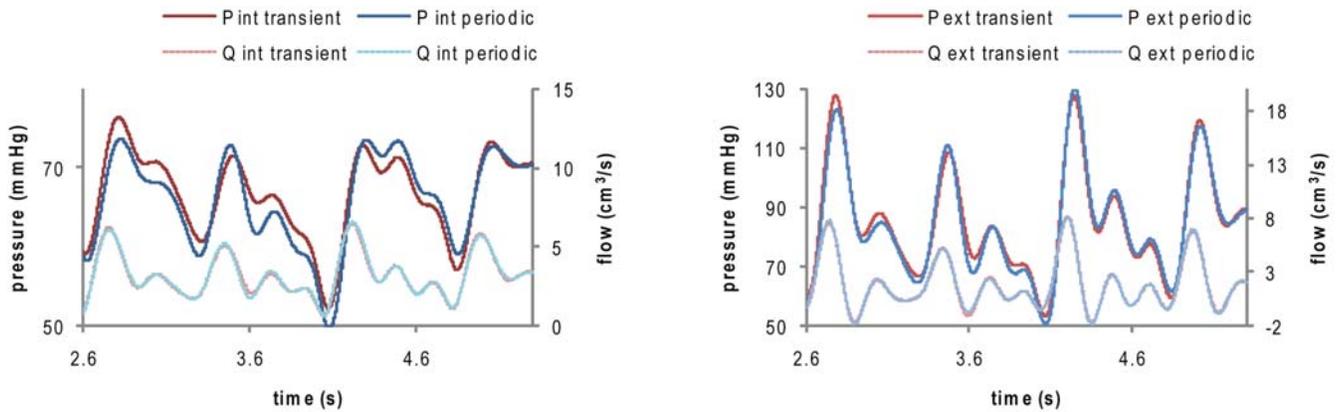

Figure 18: Comparison of flow and average pressure at the outlet boundary for the periodic RCR boundary condition and the fully transient RCR boundary condition in a carotid bifurcation with a stenosis in the internal side. Left plot: internal carotid. Right plot: external carotid.

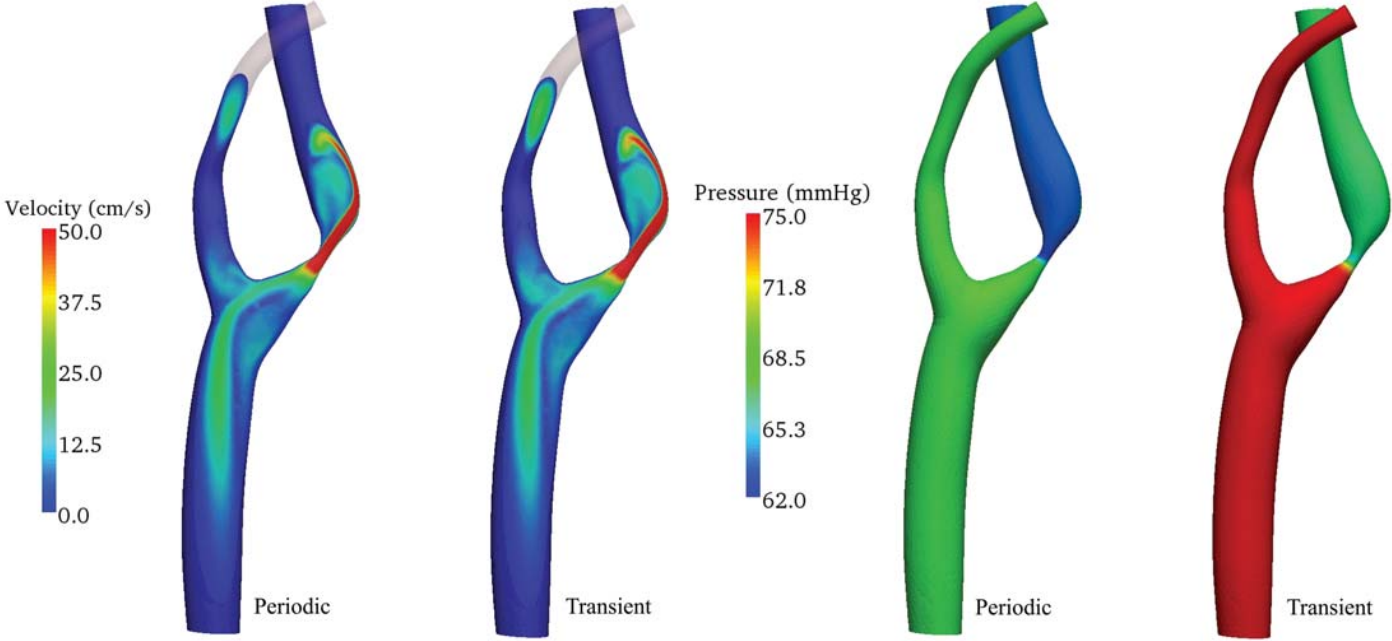

Figure 19: Comparison of velocity magnitude of a cut through the domain (left) and of pressure at the wall (right) at time 3.6 s between the periodic and fully transient RCR boundary condition simulations. The velocity contours exhibit small differences between the two boundary conditions throughout the domain. The pressure maps are significantly different between the two boundary conditions.